Genome-wide EST data mining approaches to resolving incongruence of molecular phylogenies

Yunfeng Shan[a,b]

[a]*Atlantic Bioinformatics Centre*
*6037 South Street., Halifax, Nova Scotia, Canada B3H 1S8*
[b]*Department of Biology, Life Sciences Centre, Dalhousie University*
*1355 Oxford Street, Halifax, Nova Scotia, Canada B3H 4J1*

_______________________________________________________________

**Abstract**

36 single genes of six plants inferred 18 unique trees using maximum parsimony. Such incongruence is an important issue and how to reconstruct the congruent tree still is one of the most challenges in molecular phylogenetics. For resolving this problem, a genome-wide EST data mining approach was systematically investigated by retrieving a large size of EST data of 144 shared genes of six green plants from GenBank. The results show that the concatenated alignments approach overcame incongruence among single-gene phylogenies and successfully reconstructed the congruent tree of six species with 100% jackknife support across each branch when 144 genes was used. Jackknife supports of correct branches increased with number of genes linearly, but those of wrong branches also increased linearly. For inferring the congruent tree, the minimum 30 genes were required. This approach may provide potential power in resolving conflictions of phylogenies.

_______________________________________________________________

**Introduction**

It is well known that different single genes often reconstructed different phylogenetic trees. Recent research shows that this problem still exists [1-3]. Such incongruence is caused by many reasons such as insufficient number of informative sites, lateral gene transfer, unrecognized paralogy and variable evolutionary rates of different genes [3-4]. The comparative genomics results suggest that the simple notion of a single tree of life that would accurately and definitely depict the evolution of all life forms is gone forever [5]. Other methods such as gene content and order [6-9], occurrence of orthologs and folds [10-11] throughout the whole genome were reported. However, no consistent results were obtained yet.

_______________________________

*Current Address: Molecular Genetics Laboratory, Agriculture and Agri-Food Canada, Potato Research Centre, 850 Lincoln Rd., Fredericton, NB, Canada, E3B 4Z7. Email: yshan@dal.ca

Multiple-gene approaches have been studied [12-13]. Kluge believed that phylogenetic analysis should always be performed using all the evidences [14], but Miyamoto and Fitch argued that partitions (including genes) should not be combined [15]. Concatenated alignments of a couple of genes improved supports [3-16].   Concatenating alignments into one from genome-scale 106 genes based on 7 yeast genomes [17] for phylogenetic analysis proclaimed ending incongruence [18].   However, the tree is doubted whether it is a 'true tree' [19].   Generally, determining the phylogeny of microbes was difficult due to the lack of discernible morphological characters of microbes [20]. Minimum evolution (ME) algorithm showed the different tree from maximum likelihood (ML) and maximum parsimony (MP) for the same dataset when base biases were not adjusted [21]. Therefore, this critical problem must be clarified for this approach further.   Actually, there are organisms of which the phylogeny is firmly established by fossil records and morphological characteristics [3].   If this set of organisms is used, it becomes possible to determine how reliable an approach is for reconstructing a congruent tree and overcoming the incongruence. In this study, such well-known green plants were selected. Five commonly used methods: ML, ME, neighbour-joining with unweighted least squares (NJUW), neighbour-joining with absolute difference (NJAD) and MP were used to compare consistence among methods.

There are some evident limitations for the genome-scale approach [17]. Especially, all of concatenated alignments must include exactly the same set of taxa.   This requirement limits many species representations because only dozens of species have completed genome sequence data up to now. Of them, more are microbes such as yeasts, but fewer are higher animals and plants such as rice.   In the predictable future, obtaining completed genome sequence data of all main species has to take a long time, especially for complex higher plants and animals, which have much longer genome sequences. Data mining is a powerful tool for retrieving data and is widely used in a variety of areas. An alternate solution is to retrieve enough size of EST sequence data of shared genes from publicly available sequence databases such as Genbank using data mining approach because GenBank includes many more species that have EST sequence data than the number of species that have completed sequencing genomes. The GenBank size is huge and is also growing fast. GenBank and its collaborating databases, EMBL and DDBJ, have reached a milestone of 100 billion bases from over 165,000 organisms by August 22, 2005.

**Results and Discussion**

*Incongruence among single-gene phylogenies*

Single-gene trees showed widely incongruence. 36 genes reconstructed 21 unique rooted trees using ME, 20 using NJUW, and 18 using MP (Fig. 1a-u). 8 of 36 genes reconstructed different trees using different methods (ME, NJUW, and MP). These results show that incongruence in molecular phylogenies might be resulted by not only different genes but also different methods. These 36 genes were retrieved from the EST database of GenBank through the algorithms of maximum retrieval of shared genes of six species without any consideration of their function, origin and distribution, so their phylogenetic performance is unknown and has not been previously assessed by any approach, either.

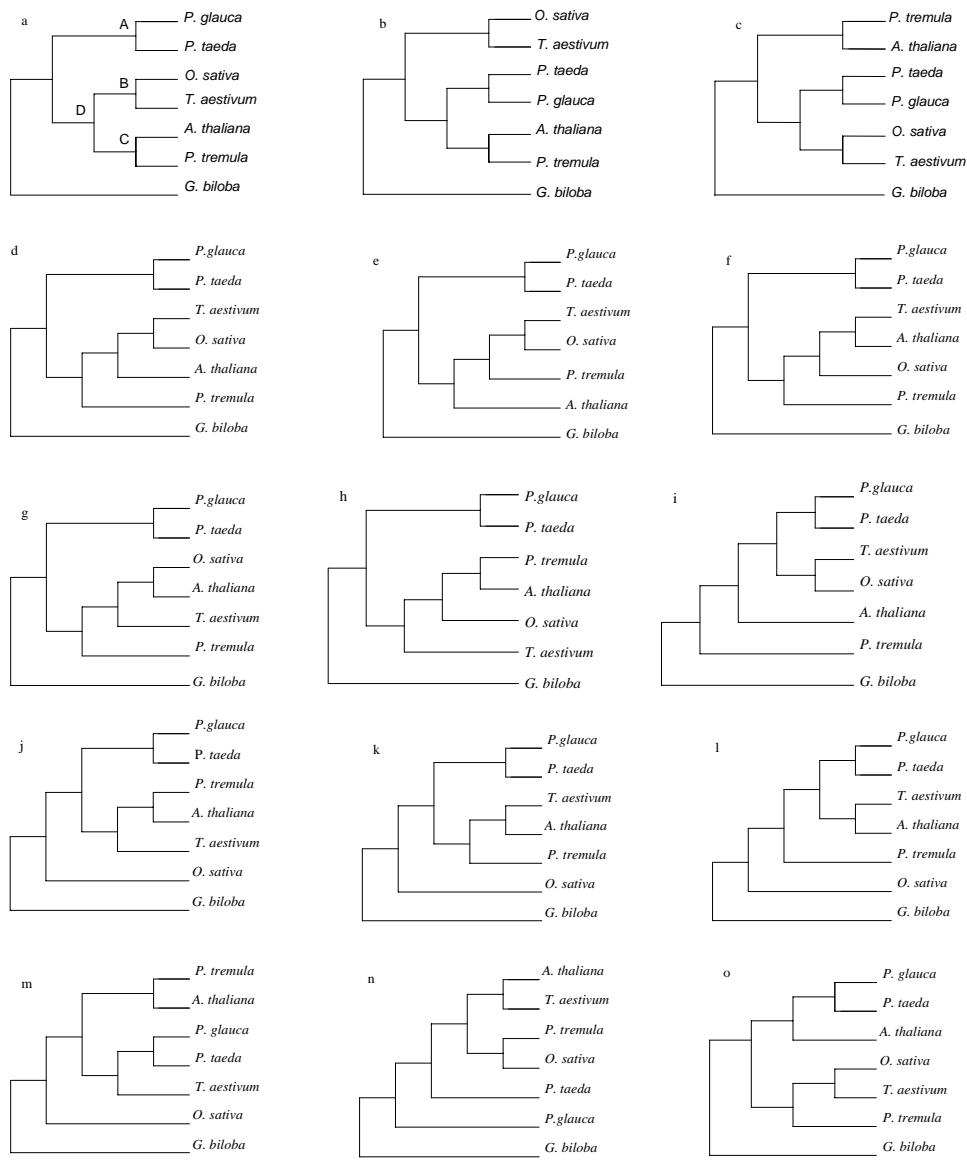

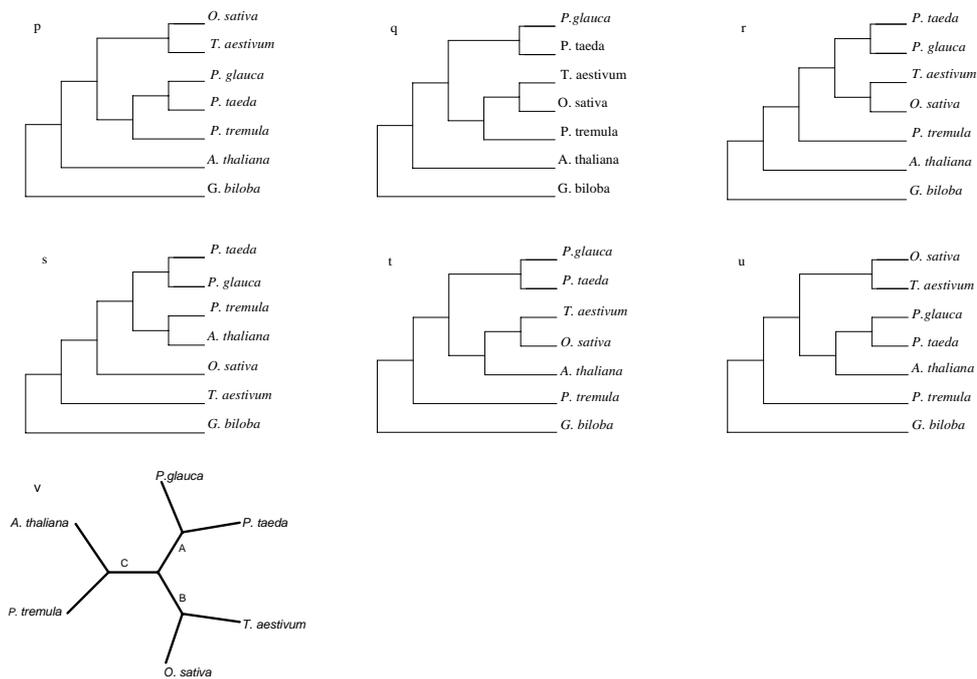

Fig. 1. Jackknife consensus trees obtained from the individual gene phylogenetic analyses of 36 shared genes with a 50% majority rule (support values not shown here). *G. biloba* was specified as the outgroup. Tree v is a unrooted tree to indicate branches (A-C), corresponding to tree a. Gene sequence datasets gave rise to each tree were not shown here, but available upon request from authors.

### *The congruent tree*

Fig. 2 and 3 show the rooted tree and the unrooted tree of the concatenated alignments of 36 genes, and Fig. 4 shows the unrooted tree of the concatenated alignments of 144 genes inferred by ML, ME, NJUW, NJAD and MP methods. The topologies of these rooted or unrooted trees all show a single tree and are consistent with those based on traditional fossil, morphological and anatomical phylogenetic analysis [22-25]. Therefore, these phylogenetic trees (Fig. 2-4) are considered as a congruent tree. The concatenation of 106 genes recovered a single species tree of 7 yeasts with maximum support of all branches [17]. However, ME tree [21] was different from the tree of MP and ML when base biases were not adjusted [17, 21]. Rokas *et al.* also observed topological differences between the tree obtained from a study on 75 yeast species but only eight commonly sequenced genes and their tree from 106 genes of seven species [17, 26]. Although Soltis *et al.* doubted whether the tree is a 'true tree', they advocated the use of multiple-gene or genome-scale approaches [19]. 106 genes of yeasts reach the genome-scale for microbes that have total hundreds of genes. 144 genes can not be said to be genome-scale for higher animals and plants that hold thousands of genes. However, the results of this study strongly show that the genome-wide data mining approach is valid to overcome incongruence for not only microbes, but also higher animals and plants regardless of the phylogenetic analysis methods, and divergence of taxa.

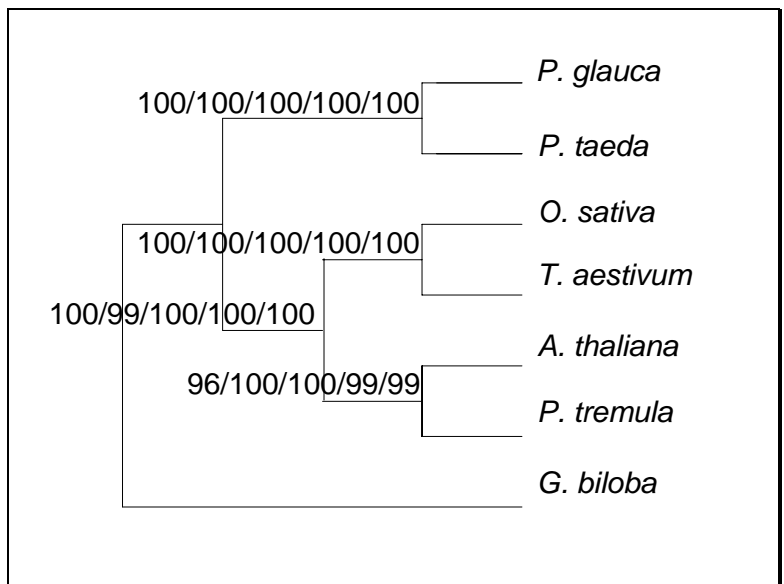

Fig. 2. The rooted tree of the concatenated alignments of 36 genes. *G. biloba* was specified as the outgroup. Numbers above branches indicate supports (ML/ME/NJUW/NJAD/MP, bootstrap for NJAD and jackknife for others).

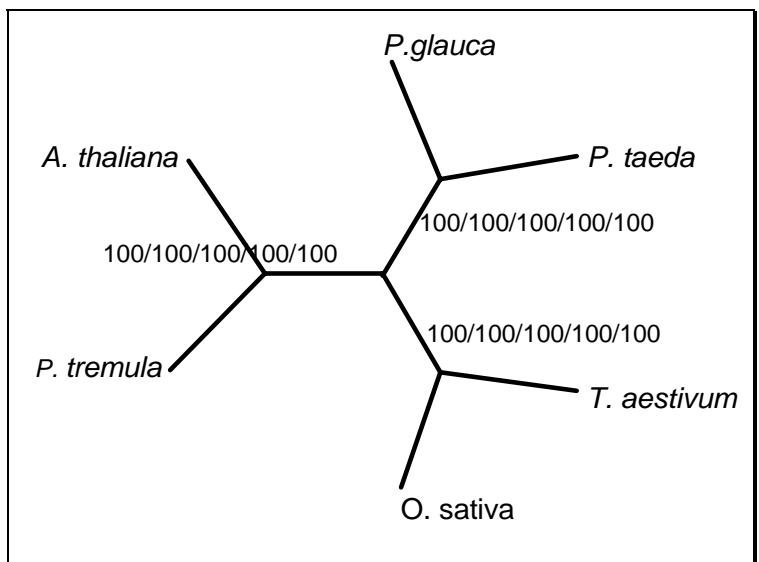

Fig. 4. The unrooted tree inferred from analysis of the concatenated alignments of 144 genes. Support is given beside branches (ML/ME/NJUW/NJAD/MP, bootstrap for NJAD and jackknife for others).

Jackknife support values for all branches of the rooted tree based on the concatenated alignments of 36 single genes of 6 species were 100% except for branch C. The support for branch C was 99% for NJAD or MP, or 96% for ML (Fig.1a and 2). Branch C was the weakest branch for this tree. Similarly, support values for all branches of the unrooted tree based on 36 single genes were 100% except that for branch C was 95%, 98%, 97%, 95% or 79% for ML, ME, NJUW, NJAD or MP, respectively (Fig. 1v and Fig. 3). Supports for all branches of the unrooted tree based on 144 genes were 100% (Fig. 4). These results suggested that more genes got greater support values. 100% support for the congruent tree from the concatenated alignments of 144 single genes regardless of the phylogenetic methods, even the simplest method such as NJAD, demonstrated power of large datasets by means of the genome-wide data mining approach in resolving the incongruence. Multiple-gene approaches such as three combined genes of angiosperms [19], four combined proteins of eukaryotes [16], six and nine genes of flowering plants [27], 23 proteins of bacteria, archea and eucarya [28], 63 genes of baculovirus [7] and 106 concatenated genes of yeasts [17] strengthened supports and improved the consistence of phylogenies. Those previous results were consistent with the result in this study.

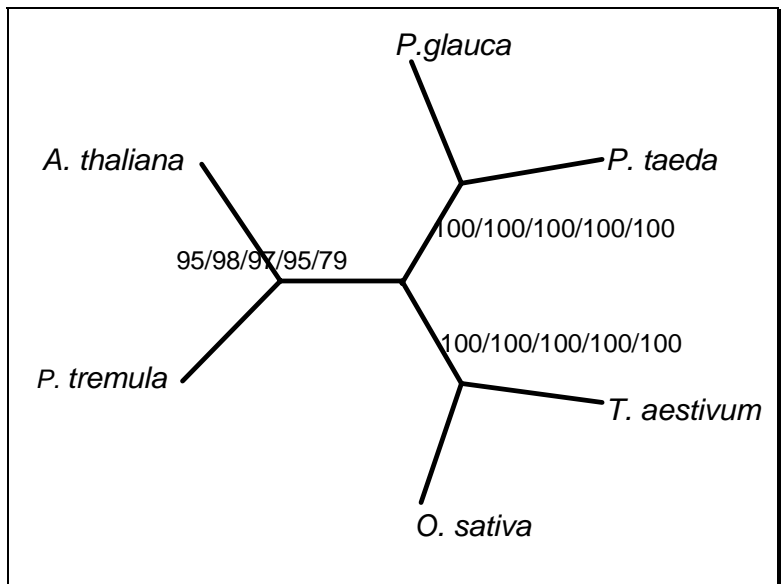

Fig. 3. The unrooted tree inferred from analysis of the concatenated alignments of 36 genes. Support is given beside each branch (ML/ME/NJUW/NJAD/MP, bootstrap for NJAD and jackknife for others).

*Minimum number of genes required to reconstruct the congruent tree*

From the dataset of 36 genes of six species, 5, 10, or 15 genes were randomly re-sampled, then one gene was added each time till 36 genes using a random number generator to determine the minimum number of genes required to recover the congruent tree. Ten replicates were used. 100% of ten replicates of 30-gene concatenated alignments got the congruent tree with an average jackknife support value of at least 85% across all branches. 80% of ten replicates of 20-gene or 25-gene concatenated alignments achieved the congruent tree with an average support of at least 70%. Only 50%, 40%, or 60% of ten replicates for 5, 10, or 15-gene concatenated alignments achieved the congruent tree with an average support of at least 70%. These results show that the number of genes sufficient to reconstruct the congruent tree with strong support across all branches was about 30 in the current case. Rokas *et al.* [17] proposed the number of genes sufficient to support all branches of the species tree ranged from a minimum of 8 to 20 based on the 106 concatenated genes of yeasts [17]. Multiple-gene approach using a small number of multiple genes can improve support, but is not sufficient to reconstruct the congruent tree. How to obtain and retrieve enough shared gene sequence data for a large size of taxa is a great challenge for wide use of the genome-wide approach.

*Jackknife supports and the number of genes*

The strongest branch often keeps 100% support even based on a single gene. The branch with the weakest support is the most susceptible but critic part for reconstructing a tree and often causes topology shift of trees, for example, branch C in this study (Fig. 1a and 2). The jackknife support values of branch C (correct branch) increased with the numbers of genes linearly (r = 0.2385**, n = 158) (Fig. 5). On the other hand, the supports of wrong branches (alternative branches of branch C) did also increase with the number of genes linearly (Fig. 6) (r = 0.4628*, n = 22). 100% support did appear in wrong branches. In a word, the genome-scale concatenation of genes increases the support regardless of correct or wrong branches. Systematic errors might accumulate with concatenating of multiple genes[21].

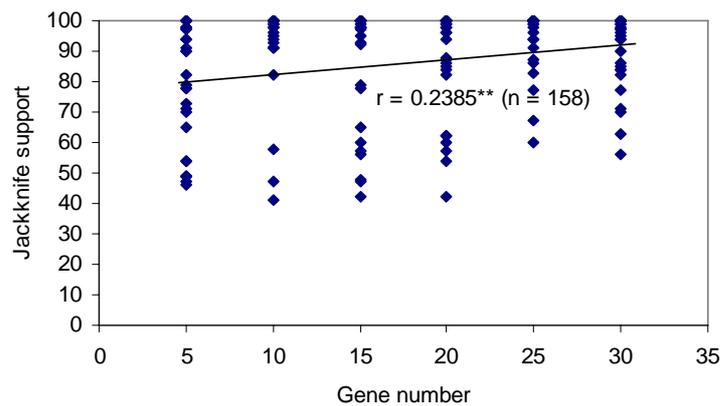

Fig. 5.    The relationship between jackknife support values of the correct branch (branch C) of phylogenetic trees and the number of genes. ** indicates the statistical significant level at 0.01.

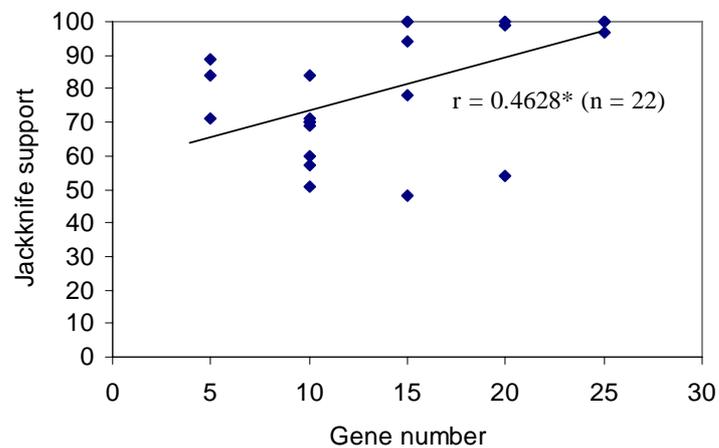

Fig. 6. The relationship of jackknife supports of wrong branches (alternative branches of branch C) and the number of genes. * indicates the statistical significant level at 0.05.

The bootstrap or jackknife method is used as a representation of confidence in phylogenetic topologies.   However, 100% support does not mean that the branch is 100% correct. 100% support may occur in an alternative branch [21]. High bootstrap support does not necessarily signify 'the truth' [19]. Even random data can yield high bootstrap support. The bootstrap or jackknife support is not suitable for genome-scale approach. A more proper index of confidence for multiple genes approach is necessary to be developed.

*Precisions of phylogenetic trees and the number of genes*

Precision is the percentage calculated by the number of correct trees divided by the total number of trees. The precisions of phylogenetic trees significantly increased with the number of genes linearly when the number of genes increased from 5 to 30 (r = 0.8039***, n = 18) (Fig. 7). The precisions to reconstruct the 'true tree' were only 30-60% when the number of genes was 5, 10 or 15. When 30 genes were used, 100% precision was observed. High precision is the appropriate candidate criteria as evidence of getting the 'true tree'.

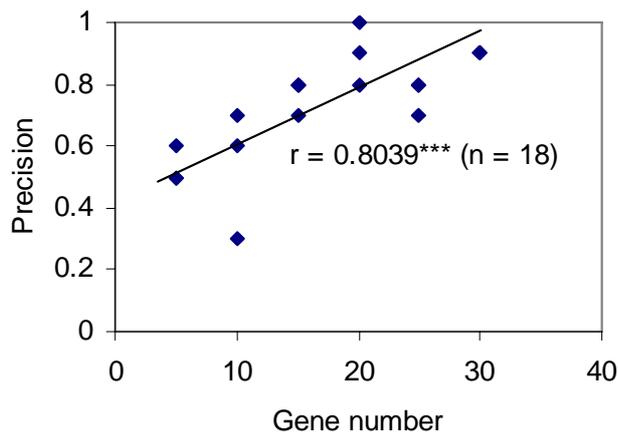

Fig. 7. The relationship between precisions of phylogenetic trees and the numbers of genes. *** indicates the statistical significant level at 0.001.

*Effects of gap handling*

Although many researchers have proposed different methods for handling gaps [29-32], no consistent algorithms exist. Gaps and uncertain positions of alignments have been sometimes handled in two extreme ways, i.e., either totally removed or totally included. The former may result in loss of some informative nucleotides or amino acids. For the latter, uncertain positions of alignments often result in uncertain phylogenies and incongruence of phylogenetic trees. In this study, when phylogenetic trees were inferred using neighbour-joining method [33] including positions with gaps [24], they were identical with the rooted tree of 36 genes (Fig. 2) and the unrooted tree of 144 genes (Fig. 4). The distances were calculated in percent divergence between all pairs of the sequence from a multiple alignment. However, branch C was not consistent between the rooted tree and the unrooted of 36 genes of six species. These results show that gaps in alignments resulted in incongruence of phylogenies. They also suggested that concatenated alignments from more genes resulted in more consistent trees and the sufficient number of genes can overcome incongruence caused by gaps. A proper algorithm to treat the positions with gaps may expect to reduce the minimum requirement of genes to recover a 'true tree'.

MP implements gaps as the fifth base for nucleotides in PAUP*4.0b10 [25]. The concatenated alignments of 36 or 144 genes including positions with gaps did not get the 'true tree': either the rooted or the unrooted. These results show that gaps caused incongruence. The gap treatment had greater impacts on the phylogenetic analysis of MP method than the distance approach. Therefore, proper gap removal is necessary and how to treat gaps still is a challenge to phylogeny inferences even using genome-scale approaches.

*Performance comparison of commonly used methods*

ML usually performs well, but is a very computing intensive method. It may not be suitable to the genome-scale approach due to very intensive computing for a large size of sequences and taxa. For general case, it is thought that the problem of ML phylogeny is NP-hard [34]. ML using branch and bound search took 19 days to complete the phylogentic analysis of the concatenated alignments of just 36 single genes of six species on a Macintosh PowerPC computer with 1.2 GHz CPU. We had to give up ML computation using branch and bound search for the dataset of 144 genes. MP took less than 1 minute for the same dataset of the concatenated alignments of 36 single genes of six species and six minutes for the dataset of the concatenated alignments of 144 single genes of six species. NJUW and ME completed computation within 5 seconds and 30 seconds, correspondingly. Powerful tree-building algorithms, such as ML, tend to rapidly become prohibitively expensive for computing time with the increase in the number of analyzed species of sequenced genomes [5]. A parallel version of high performance computing for ML or new approach will be necessary for the genome-scale approach.

**Methods**

Nucleotide sequences were retrieved from the public EST database of GenBank. Homologous genes were identified by BLASTN v2.2.6 with the highest BLASTN score hit (e-value <0.0009) available. Six species included two gymnosperms: *Picea glauca* and *Pinus taeda,* two monocots: *Oryza sativa* and *Triticum aestivum,* and two eudicots: *Populus tremula* and *Arabidopsis thaliana*. *Ginkgo biloba* was specified as the outgroup. These species were selected because their phylogeny is firmly established by fossil records and morphological characteristics [22-25]. 144 shared genes of six species except for *G. biloba* and 36 shared genes of all six species were retrieved.

Single genes were separately aligned using Clustalx with default settings [24]. All gene alignments were edited to simply exclude positions with gaps for further analysis except others specially stated, then concatenated into one large alignment for further phylogenetic analysis.

PAUP*4.0b10 [25] was used for tree inference. Each nucleotide dataset was analyzed under the optimality criteria of maximum likelihood, distance, which included neighbour-joining with unweighted least squares and minimum evolution, and maximum parsimony. The ML analyses were conducted assuming the ti/tv ratio was unequal and estimated for a nucleotide substitution model. The NJUW and ME analyses were performed assuming the HKY85 model of nucleotide substitution. The MP analyses were performed with unweighted parsimony. The jackknife consensus tree was searched using the branch-and-bound algorithm for MP, the full heuristic search for ML, NJUW and ME. Support for each branch was tested with the bootstrap/jackknife analysis. 100 replicates were used for ML, and 1000 for NJUW, ME and MP, respectively. NJAD trees were calculated using Clustalx by default settings [24], which bootstrap replicates were 1000. The correlation analyses were performed using the SAS system for Windows V8. Trees and sequence datasets are available from the authors upon request.

## Acknowledgements

Authors are grateful to the GenBank for access to the sequence database. Authors also thank Joy K. Roy for setting up this research before he left Department of Biology, Life Sciences Centre, Dalhousie University and discussed a couple of preliminary algorithms.